\documentclass[
    twocolumn,
	prd,
	amssymb,
	preprintnumbers,
	secnumarabic,
	nofootinbib,
	superscriptaddress]{revtex4-1}
	
	\pdfoutput=1

\usepackage{graphicx}
\usepackage{enumitem}
\usepackage{latexsym}
\usepackage{amsfonts}
\usepackage{amssymb}
\usepackage{color}
\usepackage{amsmath}
\usepackage{slashed}
\usepackage{dcolumn}
\usepackage{verbatim}
\usepackage{float}
\usepackage{multirow}
\usepackage{xspace}
\usepackage[normalem]{ulem}
\usepackage[
pdfauthor={Nirmal Raj}]{hyperref}

\newcommand{\ra}{\rightarrow}

\newcommand{\acro}[1]{\textsc{\MakeLowercase{#1}}} 
\newcommand{\osn}{\oldstylenums}
 
\newcommand{\beq}{\begin{equation}}
\newcommand{\eeq}{\end{equation}}
\newcommand{\bea}{\begin{eqnarray}}
\newcommand{\eea}{\end{eqnarray}}

\newcommand{\DLensSource}{D_{\rm LS}}
\newcommand{\DLens}{D_{\rm L}}
\newcommand{\DSource}{D_{\rm S}}
\newcommand{\uT}{u_{1.34}}

\newcommand{\rE}{R_{\rm E}}
\newcommand{\tE}{t_{\rm E}}
\newcommand{\vE}{v_{\rm E}}
\newcommand{\rholens}{\rho_{\rm lens}}

\newcommand{\rmax}{r_{90}}

\newcommand{\Rstar}{R_\star}
\newcommand{\rstar}{r_{\rm S}}

\allowdisplaybreaks

\begin{document}

\title{Subaru through a different lens: microlensing by extended dark matter structures
}
\author{Djuna Croon} \email{dcroon@triumf.ca}
\author{David McKeen}
\email{mckeen@triumf.ca}
\author{Nirmal Raj} \email{nraj@triumf.ca}
\affiliation{TRIUMF, 4004 Wesbrook Mall, Vancouver, BC V6T 2A3, Canada}
\author{Zihui Wang} \email{zihui.wang@nyu.edu}
\affiliation{Center for Cosmology and Particle Physics, Department of Physics,
New York University, New York, NY 10003, USA}

\date{\today}

\begin{abstract}
We investigate  gravitational microlensing signals produced by a spatially extended object transiting in front of a finite-sized source star.
The most interesting features arise for lens and source sizes comparable to the Einstein radius of the setup.
Using this information, we obtain constraints from the Subaru-\acro{hsc} survey of M31 on the dark matter populations of \acro{NFW} subhalos and boson stars of asteroid to Earth masses. These lens profiles capture the qualitative behavior of a wide range of dark matter substructures. 
We find that deviations from constraints on point-like lenses (e.g. primordial black holes and \acro{macho}s) become visible for lenses of radius 0.1~$R_\odot$ and larger,  with the upper bound on lens masses weakening with increasing lens size.
\end{abstract}

\maketitle

\section{Introduction}


Gravitational microlensing -- the transient, achromatic magnification of a star due to a transiting object -- has offered much promise in discovering dark matter lurking in macroscopic structures weighing between asteroid and solar masses.
The populations of effectively point-like lenses,
e.g. primordial black holes and \acro{macho}s,
have been constrained across a wide range of dark matter masses by surveys such as \acro{eros}/\acro{macho}~\cite{EROSMACHOCombined}, \acro{ogle}~\cite{Niikura:2019kqi}, and Subaru-\acro{HSC}~\cite{Subaru}.
The microlensing signal is appreciable when the lens comes within the Einstein radius along the line of sight between observer and source star. Lenses with spatial extent comparable to this critical distance lead to qualitatively different microlensing signals; structures with nontrivial spatial extent that have been studied include hydrogen gas clouds~\cite{WidrowHClouds} and
axion mini-clusters~\cite{Fairbairn:2017dmf,Blinov_2020},
a program recently extended by some of us to primordial subhalos and boson stars~\cite{finitelensTRIUMF}, 
and by other authors to dark \acro{macho}s~\cite{dMACHOs}.

A further complication arises when the angular extent of source stars corresponds to a distance at the lens larger than  the Einstein radius. This suppresses the magnification relative to point-like sources, as studied in detail in the case of point-like lenses~\cite{WittMao}.
The effect is applicable in particular to the Subaru-\acro{hsc} survey of M31 because of its sensitivity to small transit times and hence small Einstein radii.
It was accounted for by the collaboration by assuming that all stars in M31 have a radius of 1 $R_\odot$.
This assumption was first questioned in Ref.~\cite{Montero-Camacho:2019jte}, and later, by use of a realistic M31 stellar size distribution, shown to overestimate constraints on point-like lens populations in Ref.~\cite{SantaCruzFiniteSource}. 

\begin{figure}
    \centering
     \includegraphics[width=0.5\textwidth]{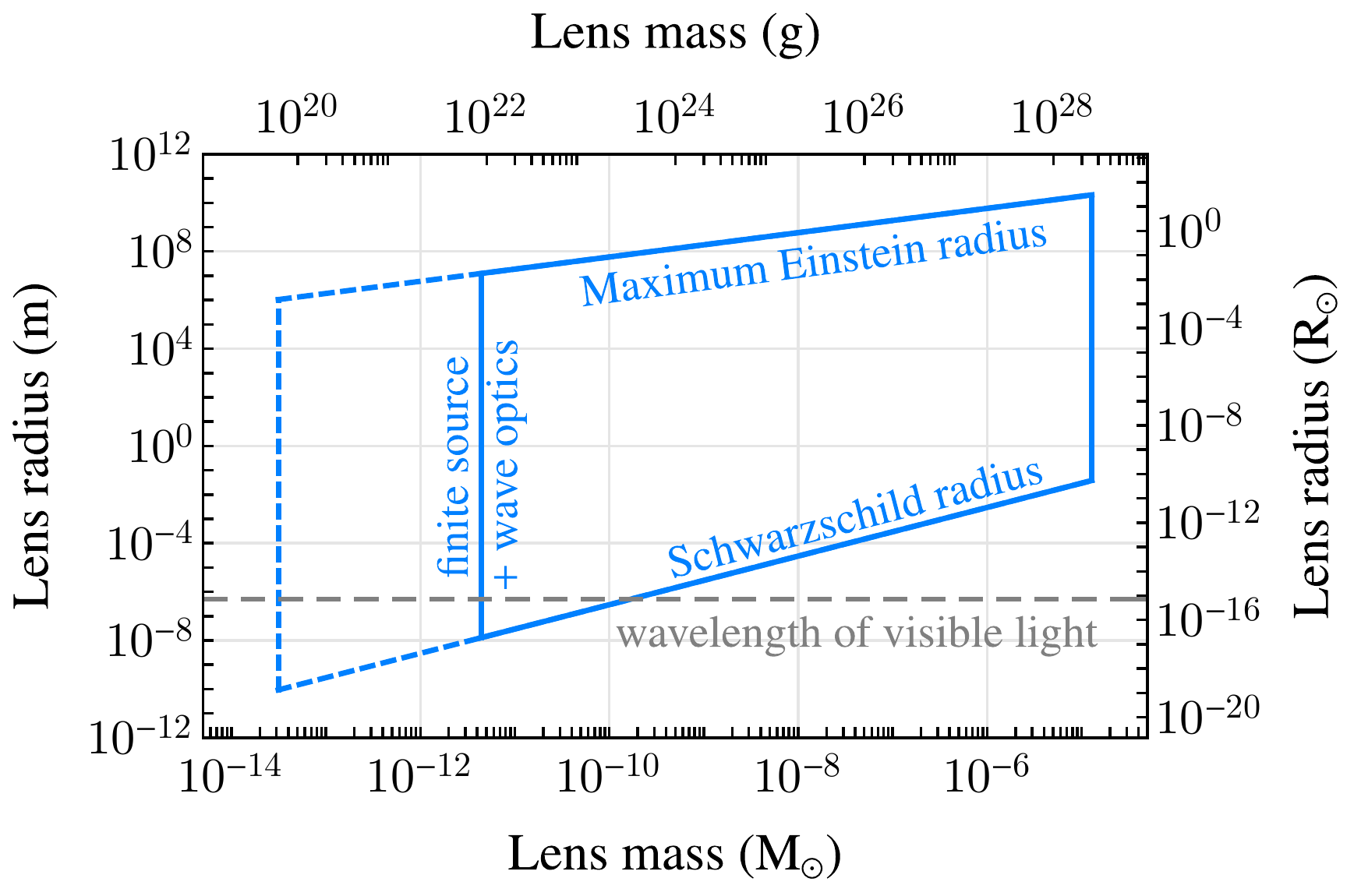}
    \caption{
    Heuristic estimate, with constant efficiencies and zero backgrounds, of the masses and sizes of dark matter structures to which the Subaru-\acro{hsc} survey is sensitive.
    Below masses of $\sim 3\times10^{-12} M_\odot$ the effects of the finiteness of the source size suppresses the microlensing magnification. 
    For point-like lenses, this is also where geometric optics breaks down and the effects of wave optics suppress the magnification~\cite{waveoptics},
    seen to occur for a lens mass corresponding to a Schwarzschild radius that is about an order of magnitude smaller than the wavelength of visible light.
    See text for further details.
     }
    \label{fig:subarutrapez}
\end{figure}

In this paper we consider microlensing constraints on extended dark matter structures using the Subaru-\acro{HSC} survey.
In Fig.~\ref{fig:subarutrapez} we show, in the space of lens size and mass, the approximate sensitivity of the survey to generic dark matter structures.
The dashed lines depict the sensitivity that might have been achieved without the effects of the sources' finite size and, in the case of point-like lenses, without the effects of wave optics.
The lowest and highest masses probed are determined respectively by the smallest and largest transit timescales to which the survey is sensitive.
For lenses much larger than the maximum Einstein radius of the setup, the lens becomes too diffuse to magnify source stars appreciably.
Moreover, lens sizes of a given mass are bounded from below by the Schwarzschild radius corresponding to that mass.

To determine the constraints on dark matter structures by the Subaru-\acro{HSC} experiment, we consider 
the microlensing signals {\em from extended sources by extended lenses}.  
This requires obtaining the magnification of images produced by such a setup, for which we outline a procedure below.
We demonstrate our procedure by studying two examples of finite-sized lens profiles that are qualitatively different, as found in our previous study: \acro{NFW} subhalos and boson stars. We expect that the constraints on a wide range of realistic dark matter structures interpolate between the constraints found in these two cases.

This paper is organized as follows.
In Sec.~\ref{sec:signals}, we describe the geometry of our set-up and outline our numerical procedure for obtaining the magnification.
We also derive the threshold impact parameter for lenses and source stars of various sizes, which determines the ``detector geometry" of microlensing.
In Sec.~\ref{sec:limits}, we count signal events and set constraints on the populations of our lens species from Subaru-\acro{hsc} observations.
Section~\ref{sec:concs} contains our conclusions.

\section{Signals}
\label{sec:signals}

The geometry along the line of sight of our setup can be found in Ref.~\cite{finitelensTRIUMF}. We denote the lens mass by $M$, and the observer-lens, observer-source, and lens-source distances by $\DLens$, $\DSource$, and $\DLensSource=\DSource-\DLens$, respectively.
In terms of these quantities the Einstein radius of a point-like lens is given by 
\beq
\rE = \sqrt{\frac{4GM}{c^2}\frac{\DLens\DLensSource}{\DSource}}=\sqrt{\frac{4GM\DSource}{c^2}  x\left(1-x\right)}~,
\label{eq:rE}
\eeq
with $x\equiv\DLens/\DSource$. $\rE$ is the closest approach to the lens of light rays from the source to the observer when the lens lies along the line of sight. It is also a useful distance scale with respect to which we normalize other distances that we introduce in the following.

\begin{figure}[b]
    \centering
     \includegraphics[width=0.4\textwidth]{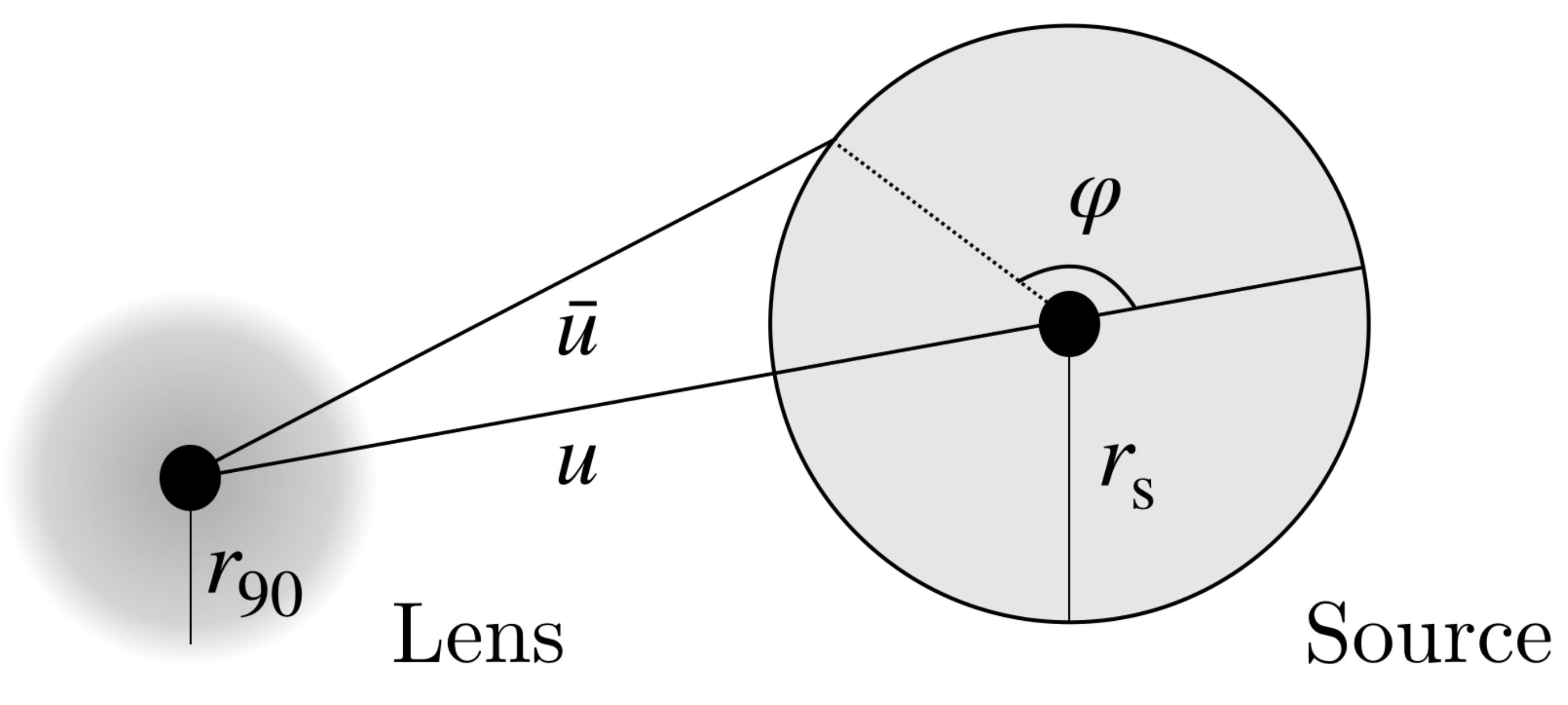}
    \caption{Geometry of our setup projected on the lens plane. 
    See Sec.~\ref{sec:signals} for further details.
     }
    \label{fig:geometrysource}
\end{figure}
The relevant distance scales along the line of sight ($\DSource$, $\DLens$) are typically much larger than those in the transverse direction (e.g., $\rE$) in the microlensing surveys we consider. This means that we can treat the lensing as occurring entirely in the transverse plane containing the lens -- for this reason it is useful to view the lensing setup projected on to this plane with all distances expressed in units of $\rE$. We have done so in Fig.~\ref{fig:geometrysource}, displaying both the finite lens and the finite source star.
In units of $\rE$, the source radius in the lens plane is $\rstar\equiv x R_\star/\rE$,
the distance from the lens center to the source center is $u$, 
and to an arbitrary point on the edge of the source is
\beq
\bar{u}(\varphi) = \sqrt{u^2 + \rstar^2 + 2 u \rstar \cos\varphi}~.
\eeq

\begin{figure*}
    \centering
      \includegraphics[width=0.48\textwidth]{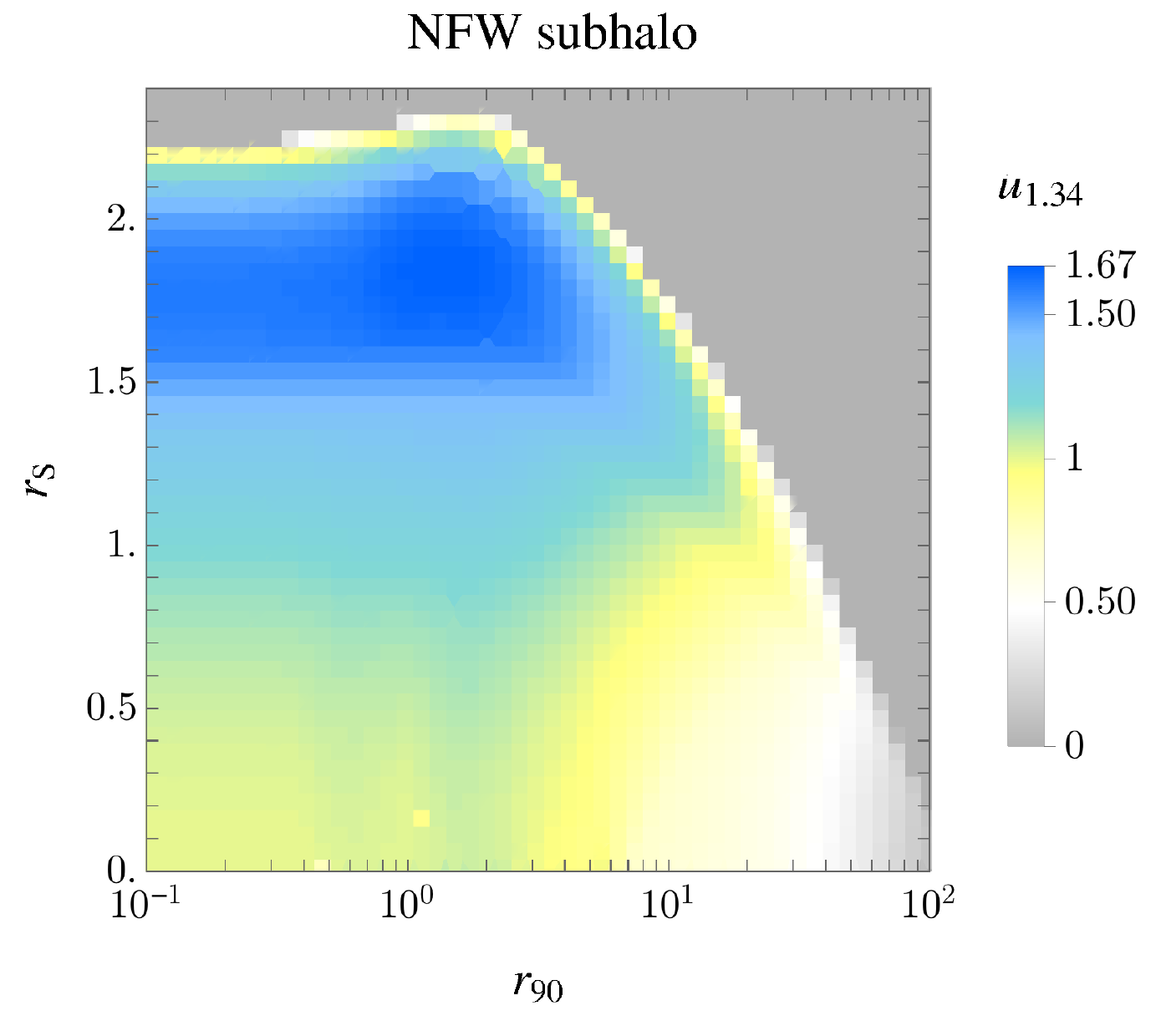}
     \includegraphics[width=0.48\textwidth]{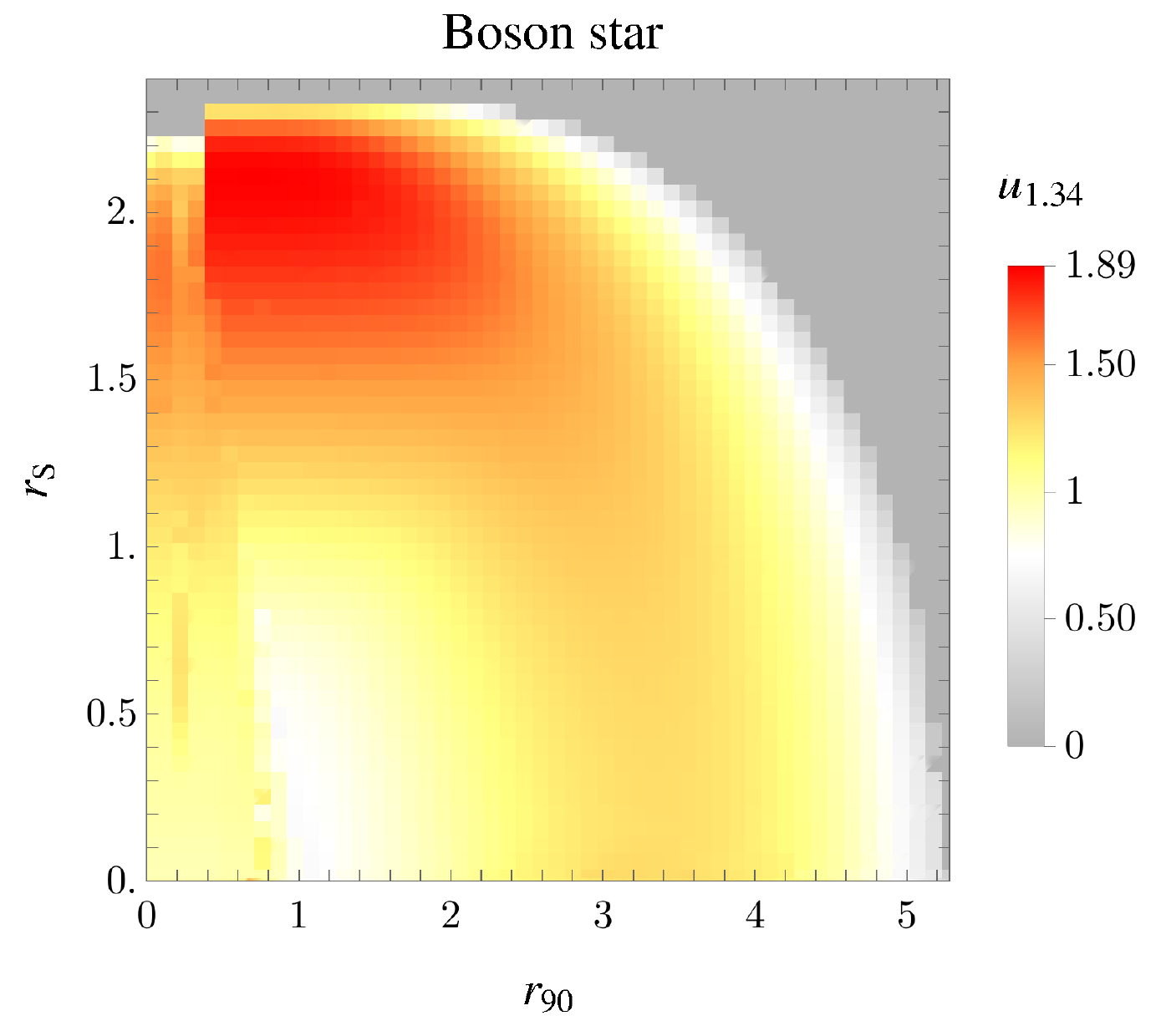}
    \caption{
    Contours of the threshold impact parameter $\uT$ in the $\rstar$--$\rmax$ (i.e. the $xR_\star/R_E$--$R_{90}/R_E$) plane for 
    \acro{nfw} subhalos ({\bf \em left}) and boson stars ({\bf \em right}).
    These provide information on the volume of space across which lens transits are counted as events. 
    Further details in Secs.~\ref{sec:signals} and \ref{sec:limits}. 
    }
    \label{fig:uTcontours}
\end{figure*}

One can then write down the lensing equation, describing the trajectory of light rays after passing the lens plane, for every infinitesimal point on the edge of the source:
\beq \label{eq:lensingeq}
\bar{u}(\varphi) = t(\varphi) - \frac{m(t(\varphi))}{t(\varphi)}~.
\eeq
Solving this yields the positions of (infinitesimal) images at $t_i(\bar{u}(\varphi))$ with $i$ labeling the, in general, multiple solutions.
As described in Ref.~\cite{finitelensTRIUMF}, $m(t)$ is the mass profile, i.e. the distribution of the lens mass projected on to the lens plane.
We refer the reader to Ref.~\cite{finitelensTRIUMF} for derivations of the mass profiles of 
\acro{NFW} subhalos, boson stars, and other lens species. In the case of an infinitesimal lens, $m(t)=1$, and the lensing equation can be solved analytically to find $|t_\pm| = |\bar{u}|/2\times \left|1\pm \sqrt{1+4/\bar{u}^2}\right|$ (note the minor typo in this solution in Ref.~\cite{Montero-Camacho:2019jte}).

Modeling the source star as having a uniform intensity in the lens plane, i.e. neglecting limb darkening, the magnification produced by an image $i$ is given by~\cite{WittMao,Montero-Camacho:2019jte}
\beq
\mu_i = \eta \frac{1}{\pi \rstar^2} \int_0^{2\pi} d\varphi \ \frac{1}{2} t^2_i (\varphi)~,
\label{eq:magfinitesource}
\eeq
where $\eta$ = sign$(dt^2_i /d\bar{u}^2|_{\varphi = \pi})$ is the ``parity" of the image. 
For the convenience of the reader, we note that the alternative formulation presented in Ref.~\cite{WittMao} should likewise contain a parity factor.
The total magnification $\mu_{\rm tot}$ is the sum of the individual $\mu_i$.\footnote{Note that our treatment ignores the effects of wave optics~\cite{waveoptics} (see also, e.g.,~\cite{Bai:2018bej,*Katz:2018zrn}) which can reduce the magnification for lens size scales (about an order of magnitude) smaller than the wavelength of light used in the survey. For our purposes, the magnification in this region of parameter space is greatly suppressed by the finite source effects that we study and so we are justified in ignoring the wave optics corrections.}

The characteristic signature of microlensing is the apparent brightening and subsequent dimming of a source star as a lens passes along the line of sight. We consider such a transit to be an event if the magnification of the source rises above threshold which, following convention, we take to be $\mu_{\rm tot}=1.34$, the magnification of a point-like source by a point-like lens when $u=1$. The time that the magnification spends above threshold is the event time, $\tE$. It is therefore convenient to define the ``threshold impact parameter" (in units of $\rE$) $\uT$ as 
\bea
\mu_{\rm tot}(u\le \uT) &\ge& 1.34~,
\label{eq:uTdef}
\eea
such that the magnification is above $1.34$ for all smaller impact parameters.

In this paper we consider two species of lenses: {\bf Navarro-Frenk-White (NFW) subhalos} that are products of hierarchical clustering, and {\bf boson stars}, structures rendered gravitationally stable by degeneracy or kinetic pressures generated by constituent scalar states.
We take these structures to be examples of the two distinct types of finite-sized lenses studied in~\cite{finitelensTRIUMF}: \acro{NFW} subhalos have a steep, peaked mass function, whereas boson stars typically follow a more uniformly distributed profile. 
As a result, boson stars may give rise to caustic crossings -- lens-source configurations where the number of images changes, implying that the magnification formally diverges -- whereas \acro{NFW} subhalos do not~\cite{finitelensTRIUMF}.

We calculate $m(t)$ for each profile as outlined in the appendix of Ref.~\cite{finitelensTRIUMF}. 
In the case of \acro{NFW} subhalos we take the lens mass profile to follow the distribution $\rho(R)\propto R^{-1}(1+R/R_S)^{-2}$ with $R$ the distance from the center of the lens and $R_S$ its scale factor. We cut off the distribution at $100R_S$ and use the radius enclosing 90\% of the total mass, $R_{90}=69R_S\equiv r_{90}\rE$, to characterize the spatial extent of the lens.\footnote{Note that due to the sharp dependence of the profile on radius for $R>R_S$, our results do not depend strongly on precisely where we cut the distribution off; other choices, unless very close to $R_S$, would lead to very similar results when expressed in terms of the characteristic physical mass and size of the lens.}
For the boson star, we solve the  Schr\"{o}dinger-Poisson equations numerically before projecting the resultant enclosed mass profile on to the plane perpendicular to the line of sight. 
Again we define $R_{90}$ as the radius which encloses $90\%$ of the total mass, and $\rmax$ as the same quantity normalized to the point-like Einstein radius. 
For both mass profiles, once we specify the lens and source sizes, we find the threshold impact parameter $\uT$ by solving the lensing equation Eq.~\eqref{eq:lensingeq} iteratively to find the impact parameter satisfying Eq.~\eqref{eq:uTdef}. 

In Fig.~\ref{fig:uTcontours} we show contours of $\uT$ in the plane of $\rstar$ vs $\rmax$.
We see that in the $\rstar \ra 0$ limit the contours follow the results we displayed in Fig.~3 of Ref.~\cite{finitelensTRIUMF}, whereas in the $\rmax \to 0$ limit the contour agrees with Refs.~\cite{WittMao,Montero-Camacho:2019jte,SantaCruzFiniteSource}.

\begin{figure*}
    \centering
    \includegraphics[width=1\textwidth]{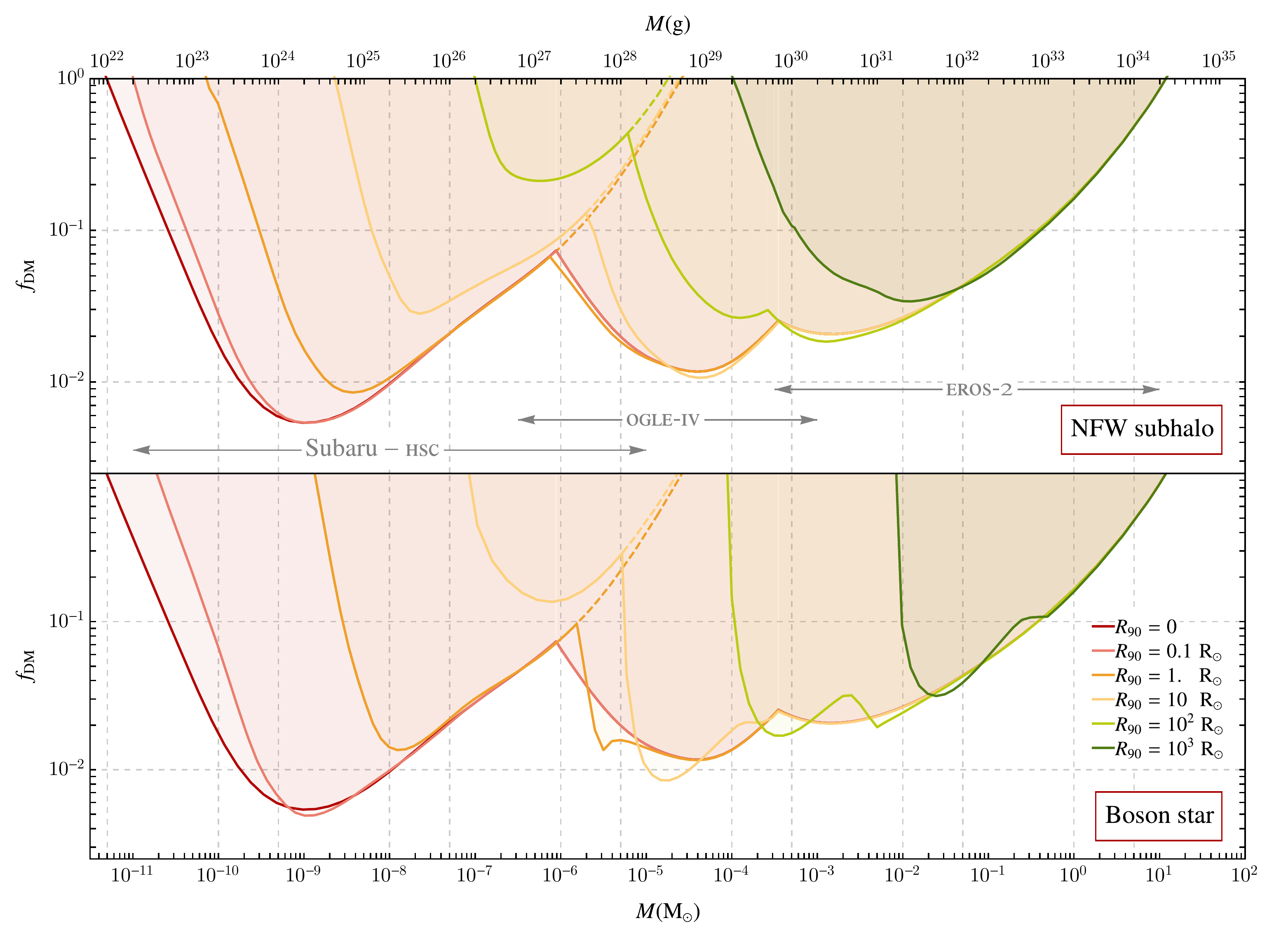}
    \caption{95\% \acro{c.l.} constraints from the Subaru-\acro{hsc} survey of M31 on the fraction of lens species making up dark matter $f_{\rm DM}$ as a function of the lens mass $M$ for \acro{nfw} subhalo ({\bf \em top}) and boson star ({\bf \em bottom}) lenses.
      These limits take into account the effect of the finite size of source stars on the microlensing signal, as characterized by the threshold impact parameter derived in Figure~\ref{fig:uTcontours},
    and the stellar size distribution of the M31 sample used in the survey.
    Also shown are constraints from the \acro{eros}-\osn{2} and \acro{ogle-iv} surveys as derived in Ref.~\cite{finitelensTRIUMF}.
    As expected, larger lenses of a given mass result in weaker limits due to smaller magnifications.
    Interestingly, the pronounced wiggles in boson star limits arising from caustic crossings seen in \acro{eros}-\osn{2} and \acro{ogle-iv} limits are absent in  Subaru-\acro{hsc} limits due to the smoothing of such features by the finite source effect.
    }
    \label{fig:fDM}
\end{figure*}

\section{Constraints}
\label{sec:limits}

Assuming that lenses have a single mass $M$, follow Maxwell-Boltzmann velocity distributions, and that their line-of-sight density is $\rholens(x) = f_{\rm DM} \rho_{\rm DM}(x)$, where $f_{\rm DM}$ is the mass fraction of lenses making up the dark matter density $\rho_{\rm DM}$,
the rate per source star is given by~\cite{Griest:1990vu}
\beq
\frac{d^2\Gamma}{dx d\tE} = \varepsilon(\tE, \Rstar) \frac{2\DSource}{v_0^2 M} \rholens(x) \vE^4(x) e^{-\vE^2(x)/v^2_0}~,
\label{eq:dGammadxdt}
\eeq
where $\vE(x) \equiv 2 \uT(x)\rE(x)/\tE$,
$\rE(x)$ is given in Eq.~\eqref{eq:rE} and
$\uT(x)$ is plotted in Fig.~\ref{fig:uTcontours} as a function of $\rmax$ and $\rstar$.
The circular speeds $v_0$ are taken from Refs.~\cite{vcircMW,vcircM31}, and are approximately 220 km/s for the Milky Way and 250 km/s for M31. 
This treatment, accounting for both the baryonic and dark matter components of the galaxies, is different from that of Ref.~\cite{Subaru} in which the circular velocity profile is computed from the enclosed mass of the dark matter halo alone, an approximation that is accurate at large distances from the galactic centers. 
Our treatment relatively mitigates the exponential suppression of the rate in Eq.~\eqref{eq:dGammadxdt}, resulting in slightly stronger bounds for large $M$, i.e. the larger circular velocity allows for lenses to transit larger Einstein radii over a shorter $t_E$.
As in Ref.~\cite{Subaru}, we adopt an \acro{nfw} profile for dark matter halo densities,
\begin{equation}
\begin{aligned}
\rho_{\rm DM} (r) &= \frac{\rho_{\rm s}}{(r/r_{\rm s})(1+r/r_{\rm s})^2}~,\\
r_{\rm MW}(x) &\equiv \sqrt{R^2_{\rm Sol} - 2 x R_{\rm Sol} \DSource \cos \ell \cos b + x^2 \DSource^2}~,\\
r_{\rm M31}(x) &\equiv \DSource(1-x)~,
\end{aligned}
\end{equation}
using parameters from Ref.~\cite{Klypin:2001xu}: $\rho_{\rm s}$ = 0.184~GeV/cm$^3$ (0.19~GeV/cm$^3$) for MW (M31),
scale radius $r_{\rm s}$ = 21.5~kpc (25~kpc) for MW (M31),
$R_{\rm Sol} = 8.5~$kpc is the distance of the Sun from the center of MW,
$\DSource = 770$~kpc is the distance to M31, and
$(\ell, b) = (121.2^0,-21.6^0)$ are the galactic coordinates of M31.

The total number of events expected is
\begin{equation}
N_{\rm events} = N_\star T_{\rm obs}\int d\tE \int d\Rstar \int^1_0 dx  \frac{d^2\Gamma}{dx d\tE}~\frac{dn}{d\Rstar},
\label{eq:Nevents}
\end{equation}
where $N_\star = 8.7 \times 10^7$ is the number of stars used in the Subaru-\acro{hsc} survey,
$T_{\rm obs}$ = 7~hr is the net observation time,
and
$dn/d\Rstar$ is the (normalized) stellar radius distribution of the source stars in M31. 
We adopt the distribution derived in Ref.~\cite{SantaCruzFiniteSource} using the Panchromatic Hubble Andromeda Treasury star catalogue~\cite{Williams_2014,Dalcanton_2012} and the \acro{MESA} Isochrones and Stellar Tracks stellar evolution package~\cite{choi2016mesa,dotter2016mesa}. 

The efficiency $\varepsilon(\tE, \Rstar)$ is shown in Fig.~19 of Ref.~\cite{Subaru} for several values of stellar luminosity (which correlates with size). To set our constraints, we approximate the efficiency as a flat 50\% for $2\,{\rm min}\leq \tE \leq 7\,{\rm hr}$.
We then locate $(f_{\rm DM}, M)$ pairs for which 
$N_{\rm events} = 4.74$, corresponding to the 95\% \acro{c.l.} Poissonian upper limit  for the one event observed at \acro{hsc}.

Our results are shown in Fig.~\ref{fig:fDM}. For completeness, we also show constraints from the \acro{eros}-\osn{2} and \acro{ogle-iv} surveys, derived with the methods outlined in Ref.~\cite{finitelensTRIUMF}. We have made the conservative assumption that the constraints cannot be combined, and provide the strongest constraint for each mass as solid curves. 
The Subaru constraints appear on the left hand side of the plots and are indicated with a dashed dividing line. 

For boson stars, the bottom panel in Fig.~\ref{fig:fDM}, we note that the features first mentioned in Ref.~\cite{finitelensTRIUMF} in the \acro{eros}-\osn{2} and \acro{ogle-iv} constraints -- which arise due to caustic crossings -- are largely smoothed out by the finite source effect. 
As a result, the primary effect of the extended lens is a loss in sensitivity to large lenses at small mass. Boson stars larger than $\sim 30 {\rm R}_\odot$ are not constrained by the Subaru-\acro{hsc} survey at any mass.

Comparing the two panels of Fig.~\ref{fig:fDM}, we note that peaked mass profiles such as \acro{NFW} subhalos lead to stronger constraints than flatter profiles such as boson stars, as expected. In particular, \acro{NFW} subhalos with $R_{90}$ up to $\sim \mathcal{O}(100){\rm R}_\odot$ have been probed by the Subaru-\acro{hsc} survey.

\section{Discussion}
\label{sec:concs}
In this work we have considered microlensing of light from the finite-sized source stars in M31 by finite dark matter structures in the M31 and Milky Way halos. The effects of these finite spatial sizes can be captured in one parameter: the threshold impact parameter $u_{1.34}$ that depends on the size of the source star as well as the spatial morphology of the dark matter structure. 
In Fig.~\ref{fig:uTcontours} we show this parameter as a function of the source size  $\rstar$ and characteristic lens radius $r_{90}$
(both normalized to the point-like Einstein radius) for two qualitatively different lens mass profiles: \acro{NFW} subhalos and boson stars. These two mass profiles interpolate between those that are  sharply peaked and those that are relatively uniform, capturing the qualitative differences in the microlensing constraints for a wide range of well-motivated dark matter substructures. Thus, our results encompass large classes of reasonable mass profiles that could be expected. Moreover, as mentioned above, because of how peaked our \acro{NFW} profile is, the details of our treatment of it, such as where we cut off the mass profile, do not strongly affect our constraints when expressed in terms of the physical lens mass and its characteristic size. We see in Fig.~\ref{fig:fDM} that nontrivial microlensing constraints exist for structures as large as $\sim 10^3 {\rm R}_\odot$, even in the case of less peaked mass functions such as obtained in boson stars. We also note that structures larger than $\sim 10^{-1} {\rm R}_\odot$ lead to modifications of the microlensing constraints compared to those of point-like lenses (as can also be observed in Fig.~\ref{fig:subarutrapez}).

We have derived conservative microlensing constraints by simply classifying a transit as an event if the total magnification rises above a threshold of $\mu_{\rm tot}=1.34$ over an appropriate time scale. In the event of the detection of a positive signal, one could use the microlensing lightcurves of finite lenses to derive further information about the properties of the lenses themselves as well as to better suppress backgrounds. For instance, particularly in the case of fairly uniform mass functions such as boson stars, the number of lensed images can change discontinuously as the lens passes along the line of sight, leading to large changes in the magnification of a source star. The interplay of this effect with the finite size of the source star could also be interesting, potentially telling us where along the line of sight the lens passed. The tools developed in this paper would allow for further study along these lines.

\appendix 

\section*{Acknowledgments}

We thank 
Sam McDermott,
Nolan Smyth, and
Sean Tulin 
for useful discussions.
The work of D.\,C., D.\,M., and N.\,R. is supported in part by the Natural Sciences and Engineering Research Council of Canada. 
T\acro{RIUMF} receives federal funding via a contribution agreement with the National Research Council Canada.

\bibliography{refs}

\end{document}